# Authentication Scheme Based on Hashchain for Space-Air-Ground Integrated Network

Caidan Zhao*‡, Mingxian Shi*, MinMin Huang*, Xiaojiang Du*†
*Dept. of Communication Engineering, Xiamen University, Xiamen, China.
†Dept. of Computer and Information Science, Temple University, USA
‡Corresponding author, Email:zcd@xmu.edu.cn

*Abstract*—With the development of artificial intelligence and self-driving, vehicular ad-hoc network (VANET) has become an irreplaceable part of the Intelligent Transportation Systems (ITSs). However, the traditional network of the ground cannot meet the requirements of transmission, processing, and storage among vehicles. Under this circumstance, integrating space and air nodes into the whole network can provide comprehensive traffic information and reduce the transmission delay. The high mobility and low latency in the Space-Air-Ground Integrated Network (SAGIN) put forward higher requirements for security issues such as identity authentication, privacy protection and data security. This paper simplifies the Blockchain and proposes an identity authentication and privacy protection scheme based on the Hashchain in the SAGIN. The scheme focuses on the characteristics of the wireless signal to identify and authenticate the nodes. The verification and backup of the records on the block are implemented with the distributed streaming platform, *Kafka* algorithm, instead of the consensus. Furthermore, this paper analyzes the security of this scheme. Afterward, the experimental results reveal the delay brought by the scheme using the simulation of SUMO, OMNeT++, and Veins.

## I. INTRODUCTION

The explosive growth in population and vehicles has brought enormous challenges to the transportation system. With the increasing number of vehicles, it is impossible to rely only on the traditional network of the ground. Therefore, Many studies have been carried out on the construction of the Space-Air-Ground Integrated Network (SAGIN) [1] [2]. [1] defines a SAGIN architecture, which can support dynamic management of resources using a Software Defined Network (SDN) controller. In [2], a space-based information system combined with the network of the ground is proposed. This system can enhance the capabilities of remote sensing, navi- gation and communication, and promote the development of real-time location services. In addition, many international companies are also conducting research and exploration. Tesla plans to launch 700 low-cost commercial satellites to provide Internet access services. Google's Loon project aims to use high-altitude balloons to provide broadband access to remote areas. Facebook is trying to deploy solar drones as well.

This work is supported by the National Natural Science Foundation of China No.91638204.

Therefore, the SAGIN will become an inevitable trend of future development.

There exist many different nodes in the SAGIN, such as vehicle nodes, roadside infrastructure, mobile terminal users, drones, airships and other stratospheric nodes, as well as high-altitude satellite nodes. Thus, it is obvious that SAGIN is confronting multiple challenges. On one hand, [3] presents a summary of the privacy and security challenges of VANET that need to be solved for the popularization of autonomous vehicles. In [4], the attacks of VANET are classified and explained along with their effects, specifically in the SAGIN. It is important to point out that the attacks of authenticity, identity, confidentiality, data integrity, and privacy must have a better solution before a wide application of the SAGIN. Another one is the stratospheric and satellite nodes also need to protect effectively. Several other papers (e.g., [28-34]) have studied related security and wireless issues. In short, problems of security in the SAGIN have drawn more and more attention.

Recently, with the characteristics of the distribution and data protection, the concept of Blockchains has been investigated intensively [5] [6]. More and more people apply them to other types of networks, especially in distributed systems, such as the Internet of Things (IoT) [7] [8] [9] [10], medical health [11], cloud computing [12], mobile commerce [13] [14], and privacy and security in distributed networks [15]. Ali Dorri et al. introduced the application of blockchains in IoT smart home security in their papers [7] [16] [17] [18] and related research. In [8], the author proposed a system based on Blockchains that can detect attacks in real time by low power consumption and future scalability of the IoT. In terms of identity authentication, privacy protection and data security, many scholars have conducted intense research [15] [19]. In [19], an out-of-band authentication scheme for IoTs based on Blockchain is proposed. The Eris Blockchain and related equipment are used to simulate the performance of the entire integrated system.

The high-speed mobility and low latency requirements of the SAGIN bring more security problems to the network. On the one hand, due to the central management, the traditional systems have many security drawbacks. Once the management nodes are attacked, the security of the entire network will face serious threats. On the other hand, based on the third-party security authentication scheme, the border crossing activities need the third-party transmission, which would cost more time.

In brief, Blockchain is distributed and decentralized. It has a good solution to the problems faced by traditional systems.

Therefore, Blockchain can be used to enhance the security of the SAGIN. In this paper, the key technologies of the Blockchain are simplified for the security problems in the SAGIN, and the cross-border identity authentication and privacy protection scheme based on the Hashchain is proposed. The contributions of this paper are as follows:

- we identify and authenticate nodes in the SAGIN by using the characteristics of the wireless signal.
- we propose a Hashchain that meets the requirements for low latency in the SAGIN.
- we establish a cross-border identity authentication and privacy protection scheme based on Hashchain.

The rest of this paper is organized as follows: Section II introduces the proposed system framework and analyzes key technologies. Security performance analysis and simulation experiments for the proposed framework are discussed in Section III. Section IV concludes this paper and presents some plans.

## II. PROPOSED FRAMEWORK AND KEY TECHNOLOGIES

This section describes the system model and certification scheme in this paper. We contrast Hashchain with the traditional structure, and describe the key technologies in detail.

### A. System Model

In the future, the vehicle can be considered as a super mobile terminal, which will be equipped with dozens or even hundreds of sensors. In the face of the numerous and complicated information resources, the vehicle need to exchange environmental information with other vehicles, roadside infrastructure, and road control centers through these sensors. Meanwhile, the dimensions and scale requirements of the situation information become more and more complicated and accurate. Since the increasing number of vehicles and the smart devices or sensors, it is not enough to simply develop the VANET. In the next part, this paper takes cross-border identity information transmission as an example, then compares and analyzes the traditional structure and the Hashchain-based structure of SAGIN.

*1) Traditional Structure:* The traditional cross-border handover authentication structure is divided into three layers, vehicle nodes, small-area management nodes (usually processed by the road side units (RSUs)), and the certificate authorities (CA). This framework is a centralized structure and is authenticated by the third party. There are two typical scenarios of cross-border identification information transmission. The first scenario is to switch from a security domain(SD) to a neighboring SD under the management of the same CA. The second scenario is shown in Fig 1(a). Vehicles travel from SD-A to SD-B, which are managed by different CAs. Therefore, the transmission of vehicle identification information requires the authentication of CA-A and CA-B constantly.

*2) Hashchain-Based Structure of SAGIN:* The cross-border identity authentication framework of the SAGIN based on Hashchain is a decentralized and distributed structure. As shown in Fig 1(b), CA is only responsible for the registration of vehicle nodes for the first time to join the SAGIN. In the Hashchain-Based framework, the management node of SD-A, called security manager (SM), adds the cross-border vehicle identity information to the blockchain, and the SM of the other SD synchronizes the account to obtain the vehicle identity information, where the vehicle is going to travel to. In addition, information can be transmitted by means of air nodes in the SAGIN, reducing the overall delay of the network.

### B. Certification scheme

In this section, a cross-border scheme based on Hashchain is proposed. The scheme uses the characteristics of the wireless signal to identify and authenticate nodes, and replaces consensus with *Kafka*, which the verification and backup of the records on the block are implemented by.

*1) Transmission Handshake:* Before entering the SAGIN, vehicles must be registered by CA. During the registration, we use the system integrated by Universal Software Radio Peripheral (USRP) or Agilent 9404 oscilloscope and computer to collect the wireless signal feature of the vehicle node as its identity fingerprint $R_v$, and records this identity information. Besides, the vehicle node record this as well. The cross domains handshake process in the SAGIN is shown in Fig 2. This algorithm to display the total scheme can be described step by step as follows:

**Step 1**: When a vehicle, $V_i$, joins to the SAGIN for the first time, it should send the beacon with an identity authentication request to the SM: $R_v$ || beacon.

SM collects the wireless signal of the vehicle, and obtains its characteristic fingerprint, $R'_v$. Comparing the collected $R'_v$ and $R_v$ sent by the vehicle, SM will realize the legality of this vehicle. If it is consistent, the vehicle is determined to be a legitimate vehicle. Otherwise, it is an illegal one.

**Step 2**: $SM_{this}$ picks up the border crossing activity of the vehicle through beacons messages sent by the legitimate vehicle, including speed, direction, and location. These border crossing activities are formed into individual transactions. Totally, SM will record $n$ transaction within a period. The transaction templates are shown in Table I.

**Step 3**: SM packages $n$ transactions into blocks, which are linked to the Hashchain based on the *Kafka* algorithm, and the block results return to the $SM_{this}$, while other SMs update their own local hyperledger at the same time. The format of Block is shown in Table II.

**Step 4**: $SM_{dest}$ decrypts the identity materials according to the private key, which would obtain that who will travel to the local area.

*2) Wireless Signal Feature:* In the process of research on individual identification, many scholars have carried out many experiments on the types of features. It is an effective method of identity authentication by the physical features [20] [21] [22]. This paper takes the individual identification of drones as an example. The [23] extracts the characteristics of the Hash fingerprint to the wireless signal, and uses the support vector data description (SVDD) algorithm to detect and identify. The average recognition rate is about 91.1%. In our framework, the function which represents the Hash fingerprint feature of wireless signal has been defined as an array consisting of 1, -1, 0, named $R_{node}$. The feature fingerprint of the wireless signal is unique and unclonable compared with the traditional public-private key scheme.

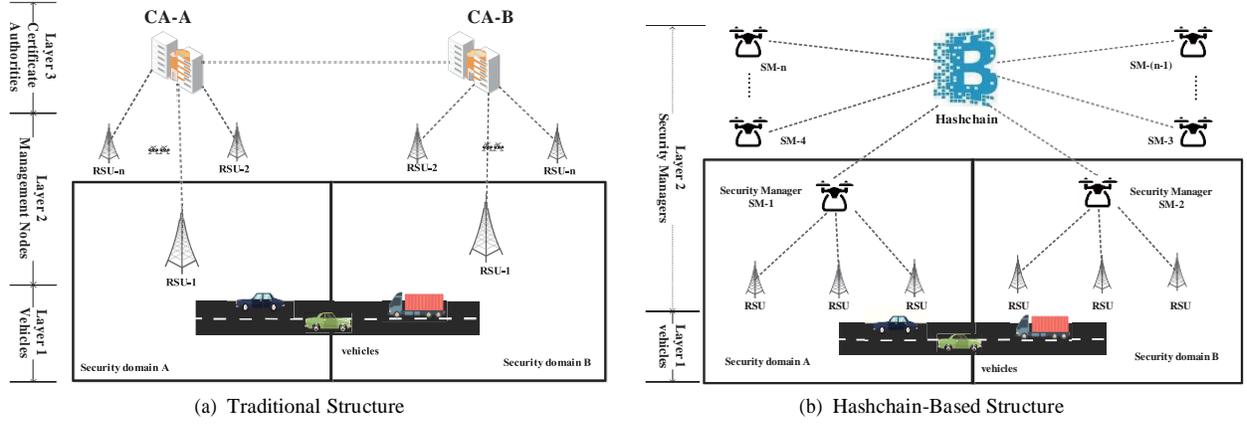

Fig. 1. Feature Extraction of different domain

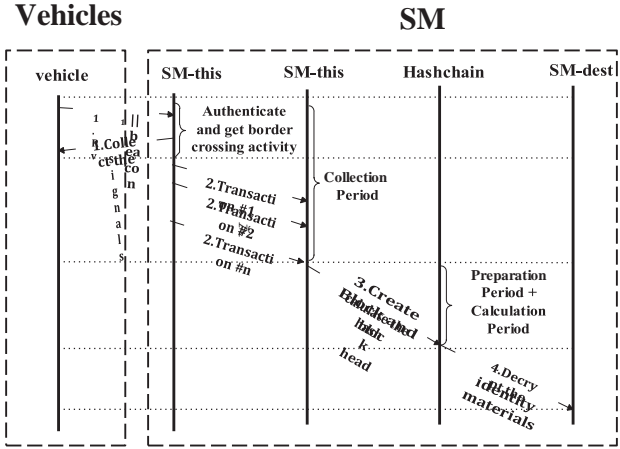

Fig. 2. Transmission Handshake

*3) Hashchain Format:* In combination with the characteristics of SAGIN transmission, a detailed presentation of the Hashchain format is defined, as shown in Table I. Transactions are developed to encapsulate identifiable materials from the source SM to destination SM. The Transaction Header contains five fields, the identity materials of source SM, $R_{this}$, the transaction number $i$, the identity of destination SM, $R_{dest}$, the vehicle identity materials including the encrypted data and the timestamp. Source and destination SMs are considered as the input and output in the traditional Blockchain system [5]. The transaction number relates to the position of this transaction in the transaction packet. Whats more, we use the public key of the destination SM for encryption, so that the other SM cannot read these materials. The destination SM uses its private key to decrypt the vehicle identity materials, who is going to this SD. In addition, current SM could package all transactions collected per unit time into blocks.

As shown in Table 2, the format of block consists of the identity of the source SM, the Previous Block Hash, the Timestamp, and Merkle Tree Root. The second field describes the hash of the previous block in the chain that directly links to the parents block. As for the Merkle Tree Root, it is obtained through all transaction in the block. Each record of the transaction records is hashed, and then combined randomly with any transaction to get a new hash value. Finally, all transaction are merged into the Merkle Tree Root, which will build up the protection of the integrity. Once any transaction is changed by someone else, the Merkle Tree Root will change consistently. Additionally, the Timestamp remains the creation time of this block.

*4) Kafka:* In the traditional blockchain, such as bitcoin, it may take almost 10 minutes to create a block. It is obvious that this delay doesn't meet the demands of time in the SAGIN. Therefore, we simplify the traditional blockchain and implement the verification and backup of the records on the block with a distributed message processing, *Kafka*, without relying on consensus.

Apache *Kafka* [24] is a distributed streaming platform. Best of all, *Kafka* has high throughput and low latency. The kafka cluster stores streams of records in categories, called topics. For each topic, the kafka cluster maintains a partitioned log, which is an ordered, immutable sequence of records. Specifically, there are two main roles, producers and consumers. Producers publish a stream of topics into the kafka cluster in this way, which can push the data as a batch to improve the efficiency. In fact, we can modify the parameters of the batch to create a new block. For example, we can set the different values of the cumulative quantity, the cumulative time interval, or the accumulated data size. Another role is consumers that can subscribe to one or more topics from the kafka cluster. In general, Producers and consumers are considered as input and output in the Hashchain system. The Hashchain-based authentication algorithm is shown in Algorithm 1.

TABLE I
FORMAT OF TRANSACTION.

| Transition Header |
| --- |
| Identity materials of source SM, $R_{this}$ |
| Number of this transaction |
| Identity materials of destination SM, $R_{dest}$ |
| Vehicle identity materials including the encrypted data. (pseudonym $V$ \|\| vehicle identity, $R_v$)$PK_{SMdest}$ |
| Timestamp |

TABLE II
FORMAT OF BLOCK.

| Block Header |
| --- |
| Identity materials of the source SM, $R_{this}$ |
| Previous Block Hash |
| Timestamp |
| Merkle Tree Root |
| Payload(Transaction) |
| Transaction N0.1, $\cdots$, Transaction No.n |

---

**Algorithm 1:** Create Block based on $Kafka$

**Input:** Information to create candidate Block Header **H**: Block Version $V_B$, Previous Block Hash $Hash_{pre}$, this identity of SM, $R_{SM-this}$, Timestamp $t_{now}$, Transactions $T_{tran} = [T_1, T_2 \ldots T_n]$, Cumulative quantity, $N$.

**Output:** the Hashed Block Header $H_{temp}$

1: Initialise bool variable $IsBlock = FALSE$;
2: Initialise the sum of transactions in this block $Sum$;
3: Initialise $result = randPerm(T_{trans})$
4: Remove $result$ from the transaction pool, $result \notin T_{trans}$;
5: **for** $Sum \leq N$ **do**
6:    Select randomly $T_i = randPerm(T_{trans})$ and $T_i \neq result$;
7:    Remove $T_i$ from the transaction pool, $T_i \notin T_{trans}$;
8:    $result = dhash(T_i, result)$;
9:    $Sum++$;
10: **end for**
11: Calculate Merkle tree root $Root_M$ basing on $result$
12: Creat the hash block heard $H_{temp}$
   $H_{temp} = V_B /\!/ Hash_{pre} /\!/ Root_M /\!/ t_{now} /\!/ R_{SM-this}$
13: Start a Zookeeper server;
14: **while** Not $IsBlock$ **do**
15:    Use $Kafka$ connect to import data to the chain;
16: **end while**
17: **return** the Hashed Block Header $H_{temp}$ and Block Number

---

## III. ANALYSIS AND SIMULATION

In this section, we evaluate the performance of our proposed framework in two major aspects. One is the security analysis of the entire framework, and the other is the delay simulation and analysis of the framework.

### A. Security Analysis

This paper takes the security of Hashchain into consideration when the Hashchain is adopted as the basic framework of the whole authentication process. The most important features of the Hashchain are: 1) distributed and 2) decentralized. These above statements mean that the Hashchain structure is regarded as a Peer-to-Peer(P2P) topology. Concretely, the following is an analysis of the security and privacy of the cross-border identity authentication scheme based on the Hashchain in the SAGIN. Table III lists some of the security requirements and corresponding solutions.

The framework leverages the effective data security features of Hashchain to ensure security in the SAGIN. Specifically, this paper analyzes some questions from the following aspects:

- Confidentiality: The identity materials need to be encrypted and transmitted by a hash function.
- Integrity: The storage of each action/data is based on the Merkle tree. Distinctly, we can easily learn that whether the data has been tampered with relying on the Merkle Tree Root. It is able to ensure the integrity of the data in this way.
- Non-repudiation: In this network, every SM will have timestamp and signature when recording. Besides, the other peers will update their local hyperledger, so that it is fairly easy to compare and verify by different nodes.

### B. Network Simulation

This paper simulates the experiments by SUMO [25], OMNeT++ [26] and Veins [27]. Here, SUMO is used to simulate road traffic, OMNeT++ implements network framework simulation, and Veins combines these two software to simulate complex communication of nodes. We mainly consider the delay of communication and authentication between the vehicles and the low-altitude management nodes.

*1) Simulation Setup:* In this simulation, the nodes communicate with each other based on IEEE $802.11p$ with the parameters as Table IV. Meantime, the simulation scenario is shown in Fig 3. The low-altitude drone is used as the source domain management node and the vehicles are generated from the four vertex positions of the route respectively, which travel from SD-A to SD-B along the road. The vehicle speed is $8.33m/s(30km/h)$, $13.89m/s(50km/h)$, $19.44m/s(70km/h)$, $27.78m/s(100m/s)$, the number of vehicles increased from 20 to 120, with a step size of 20.

*2) Result:* For the identity detection and identification, we equip the USRP and other devices to collect the wireless signals sent by the nodes in real time and then use the trained model to classify the signals. The experimental results indicate that the whole process takes about $11ms$ on average.

TABLE III
SECURITY REQUIREMENTS AND SOLUTIONS IN THE SPACE-AIR-GROUND INTEGRATED NETWORKS

| Security Requirements | Solutions |
|---|---|
| Nodes authentication | We can use the characteristics of the wireless signal to identify and authenticate nodes because of its uniqueness and unclonability. |
| Various attacks (Sybil attacks, Denial of Service attacks, Replay attacks, etc.) | Once a node launches an attack in the entire network, it will be discovered and eliminated by other nodes. A node is broken, and the data record about the node is backed up elsewhere. |
| Data security | Due to the Merkle Tree Root, the record in the Hashchain is tamper-resistant and undeniable. |
| Efficiency and low latency | Simplifying the Blockchain, we replace consensus with *Kafka* |
| Privacy protection | we put forward the method based on the pseudonym |

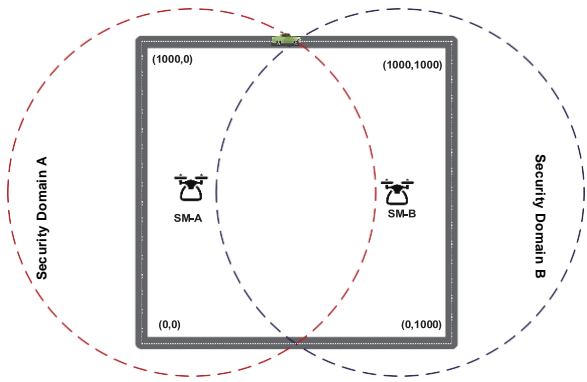

Fig. 3. Simulation scenario

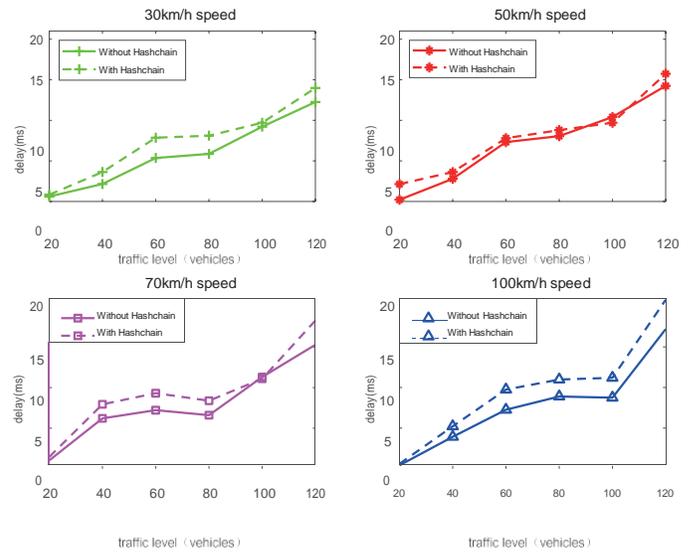

Fig. 4. The comparison between Hashchain and non-Hashchain structure about delay

TABLE IV
SIMULATION SETUP

| Parameters | Values |
|---|---|
| Frequency/($ttHz$) | 5.89 |
| IEEE 802.11p data rate/($Mbps$) | 6 |
| sensitivity in physical layer/($dBm$) | -89 |
| Number of SM | 2 |
| Total area/($km^2$) | $1 \times 1$ |
| Simulation time/($s$) | 360 |
| Vehicular density | (20-120) |
| Vehicle speed/($km/h$) | 30,50,70,100 |
| beacon size without Hashchain/($bit$) | 80 |
| Hashchain-Based beacon size/($bit$) | 720 |

In practice, a pre-trained model is used for detection and classification, so the delay is greatly reduced.

Additionally, in order to compare the changes about delays between Hashchain structure and without Hashchain, we design this simulation model. The results are shown in Fig 4.

The dotted line indicates the time cost of Hashchain-based authentication in the SAGIN, and the solid line illustrates the delay overhead without Hashchian. Also, different colors represent the results of different vehicle speeds. At the same speed, with the increment of vehicles, the average end-to-end delay increases consistently. It is obvious that the average delay required for the authentication scheme of this paper is larger than the one without Hashchain. Taking the vehicle speed of $30km/h$ as an example, the authentication delay without Hashchain is $5.9ms$, and the authentication process delay with the Hashchain is $7.36ms$. In the case where the vehicle speed is $50km/h$, the gap of delay between the two schemes is smaller, because that the speed of the vehicle matches the number of transactions. That means the delay of the entire network waiting for the transaction to form a block is shorter. From the overall situation, the average latency of authentication based on Hashchain is $8.6ms$, which is greatly reduced compared to the traditional blockchain generation time. On the other hand, further research is needed to achieve lower latency requirements.

IV. CONCLUSION

This paper proposes a robust Hashchain-based authentication scheme, which manages the data information in the network through the effective data security features of the Hashchain. At the same time, this paper illustrates security analysis and simulation results. The framework meets the security requirements in the SAGIN, where the overall authentication delay average is $8.6ms$. In future work, we will further quantitatively study the number of nodes and transactions of the Hashchain in the SAGIN, and balance the storage space and time overhead of the network. Moreover, we will compare the traditional authentication scheme with the Hashchain-based scheme which is proposed in this paper.